\documentclass[sigconf]{acmart}
\usepackage{amsmath,nccmath,tabularx} 
\usepackage{algorithm}
\usepackage{algorithmic}
\usepackage{graphicx}
\graphicspath{{images/}}
\usepackage{wrapfig,lipsum}
\usepackage{tabularx}
\usepackage{subcaption}
\usepackage{multirow}
\usepackage{hyperref}
\usepackage{makecell}
\usepackage{fixmath}
\usepackage{enumitem}

\newcommand\numberthis{\addtocounter{equation}{1}\tag{\theequation}}

\AtBeginDocument{%
  \providecommand\BibTeX{{%
    \normalfont B\kern-0.5em{\scshape i\kern-0.25em b}\kern-0.8em\TeX}}}

\copyrightyear{2022} 
\acmYear{2022} 
\setcopyright{acmlicensed}\acmConference[SIGIR '22]{Proceedings of the 45th International ACM SIGIR Conference on Research and Development in Information Retrieval}{July 11--15, 2022}{Madrid, Spain}
\acmBooktitle{Proceedings of the 45th International ACM SIGIR Conference on Research and Development in Information Retrieval (SIGIR '22), July 11--15, 2022, Madrid, Spain}
\acmPrice{15.00}
\acmDOI{10.1145/3477495.3531994}
\acmISBN{978-1-4503-8732-3/22/07}

\begin{document}
\fancyhead{}
\title{Implicit Feedback for Dense Passage Retrieval:\\A Counterfactual Approach}

\author{Shengyao Zhuang}
\affiliation{%
	\institution{The University of Queensland}
	\streetaddress{4072 St Lucia}
	\city{Brisbane}
	\state{QLD}
	\country{Australia}}
\email{s.zhuang@uq.edu.au}

\author{Hang Li}
\affiliation{%
	\institution{The University of Queensland}
	\streetaddress{4072 St Lucia}
	\city{Brisbane}
	\state{QLD}
	\country{Australia}}
\email{hang.li@uq.edu.au}
\author{Guido Zuccon}
\affiliation{%
	\institution{The University of Queensland}
	\streetaddress{4072 St Lucia}
	\city{Brisbane}
	\state{QLD}
	\country{Australia}}
\email{g.zuccon@uq.edu.au}


\begin{abstract}
In this paper we study how to effectively exploit implicit feedback in Dense Retrievers (DRs). We consider the specific case in which click data from a historic click log is available as implicit feedback. We then exploit such historic implicit interactions to improve the effectiveness of a DR. 
A key challenge that we study is the effect that biases in the click signal, such as position bias, have on the DRs. To overcome the problems associated with the presence of such bias, we propose the Counterfactual Rocchio (CoRocchio) algorithm for exploiting implicit feedback in Dense Retrievers. We demonstrate both theoretically and empirically that dense query representations learnt with CoRocchio are unbiased with respect to position bias and lead to higher retrieval effectiveness. We make available the implementations of the proposed methods and the experimental framework, along with all results at \url{https://github.com/ielab/Counterfactual-DR}.
\end{abstract}


\maketitle

\section{Introduction} \label{intro}

Dense Retrievers (DRs) are retrieval and ranking approaches where a transformer-based deep language model (e.g., BERT~\cite{kenton2019bert}) is used to separately encode queries and documents in low dimensional embeddings (dense vectors), which are then used as representation for matching via vector similarity search (inner product). 
Although generally less effective than cross-encoder approaches such as mono-BERT~\cite{nogueira2019passage}, DRs offer much lower query latency, overcoming the main barriers to adoption of cross-encoder methods (high query latency and high computational cost).


Dense retrievers are static, in the sense that after a DR has been trained, its encoding mechanism does not change, and so does not its retrieval mechanism. 
In this paper, we are interested to investigate the use of implicit feedback that is collected by a search engine, and in particular click-through data~\cite{agichtein2006improving,joachims2017accurately}, in the context of DRs. In particular, we are interested to explore how the click signal collected in historical click logs could be used to improve the effectiveness of DRs. 

\begin{figure}[t]
	\includegraphics[width=1\linewidth]{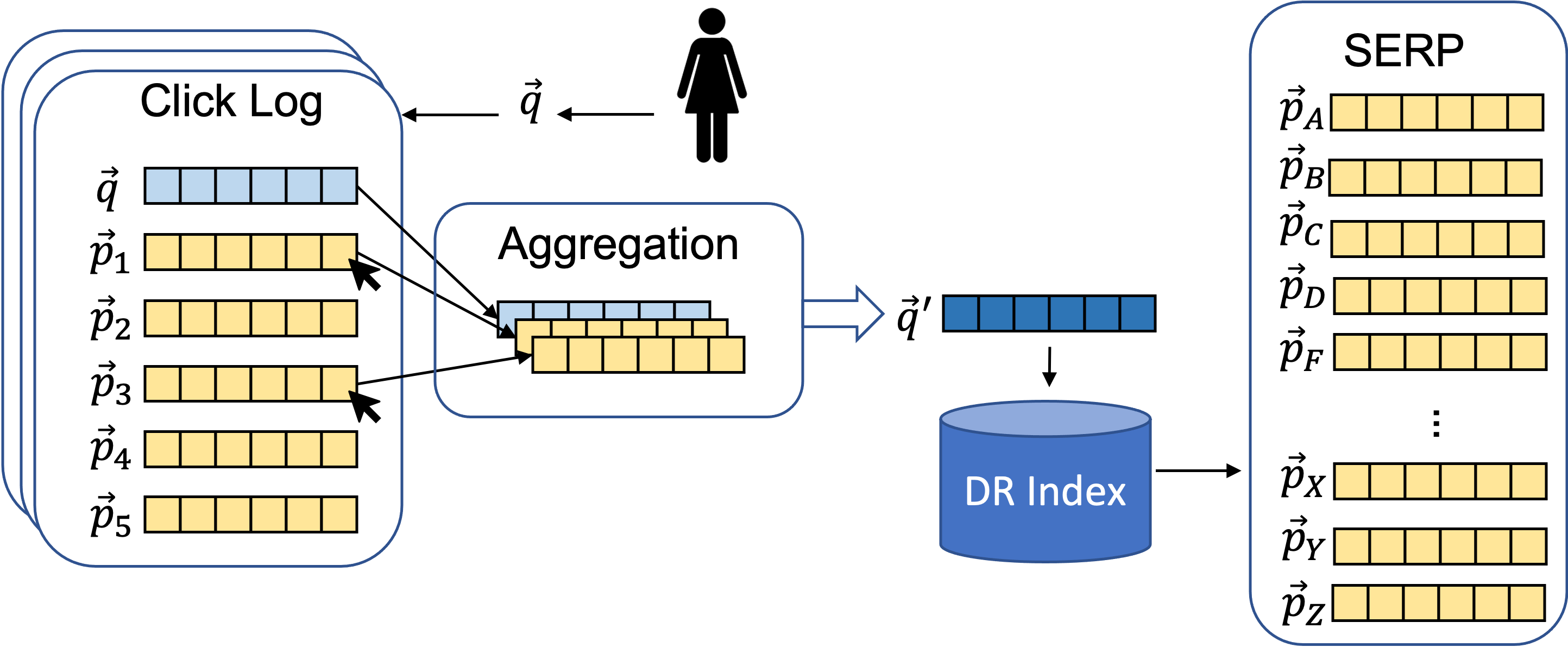}
	\caption{High level visualisation of our proposed method for exploiting implicit feedback from historic click logs to enhance dense retrievers (DRs).}
	\label{fig:implicit}
\end{figure}

The key idea we put forward in this paper is to adapt current pseudo relevance feedback (PRF) methods for DRs~\cite{li2021pseudo} to deal with implicit feedback in terms of clicks. 
A first sight, this can be achieved by simply removing the PRF assumption that the top-k search results are relevant and therefore form the relevance signal. This assumption can be replaced by the assumption that the clicked search results are what constitutes the relevance signals and can be used to enrich the query representation. This intuition is depicted in Figure~\ref{fig:implicit} where the dense representation of the query $\vec{q}$ is aggregated with the dense representation of the clicked passages $\vec{p_1}$ and $\vec{p_3}$, to create the updated query $\vec{q}'$. 
Note that a difference exists here between the traditional use of PRF and our adaptation, besides from the type of relevance signal. PRF is commonly used to execute a second round of retrieval, following on from an initial retrieval round without feedback. The methods we put forward in this paper instead tackle the first round of retrieval: the click feedback does not come from the interaction between the current user and the SERP, but from the historic click feedback the system has collected for past queries.

The adaptation of current PRF methods for DRs to dealing with the click signal, however, presents three key challenges.
	
		\textbf{Challenge 1 (datasets):} the training of DRs requires large datasets, such as MS MARCO, that contain both textual passages and relevance labels. These datasets however do not have corresponding click data to be used as implicit feedback to support our investigation\footnote{While there exist a clicklog that is associated to the MS MARCO dataset (ORCAS~\cite{craswell2020orcas})  this does not fit the needs for our study -- see section~\ref{related} for details.}. Then, how can existing large datasets for training and evaluating DRs be extended with click information? 
	
	To address this challenge, we use the evaluation practice developed in online learning to rank (LTR) and counterfactual LTR, where click models are used to simulate user interactions. This allows us to simulate clicks on passage ranking collections used to study dense retrievers.

	\textbf{Challenge 2 (click bias):} it is well known that the click signal presents a number of biases, such as position bias~\cite{yue2010beyond,joachims2017unbiased}. Then, what is the impact of this bias on the effectiveness of relevance feedback methods for DRs when clicks are used as relevance signal? How can this bias be removed from the click signal -- and does this improve search effectiveness?
	
	To investigate and address this challenge, we first study the impact of implicit feedback on the effectiveness of DRs under different user conditions and click bias. We then propose a counterfactual method for DRs that mitigates the impact of bias in the click signal. Specifically, our method relies on a historic click log to estimate the likelihood of position bias associated to a query $\vec{q}$. This is then used to adjust the weight of the dense representation associated to a feedback passage $\vec{p}$ when aggregating this signal with that from other passages and the original query $\vec{q}$ to compute the new dense query representation $\vec{q}'$. 
	

	\textbf{Challenge 3 (counterfactual DRs with unseen queries):} our counterfactual method requires that, for any given query to be scored, the historic click log contains interactions with SERPs for that query. While this is a reasonable expectation for popular queries, it is likewise reasonable to expect that at times users will enter queries that are not present in the historic click log. What strategy can then be used to adapt the proposed counterfactual method to cases in which the query has not been previously observed in the historic click log?
	
	To address this final challenge, we first identify a strategy to generate new queries not in, but yet related to\footnote{Such that the signals from the historic click log bear a valuable signal for the unseen queries.}, the historic query log we create as dataset. We then devise a simple but effective strategy to identify the portion of the signal from the historic click log to be used by our counterfactual method for the current query.
	

In summary, this paper explores a novel research direction: that of exploiting implicit signals such those in historic click logs to enhance DRs. With this respect, we show that evaluation practices developed for online and counterfactual LTR can be adopted to this context. In addition, we show that the historic click signal can be highly informative for DRs, making these methods highly effective. On the other hand, we show that the noise and bias typically found in click-through data have a significant detrimental effect. To address this issue we contribute a novel counterfactual method for DRs to exploit biased historic click signals, called Counterfactual Rocchio (CoRocchio) and investigate the key factors that influence its effectiveness, including its application to two different DRs and how it can deal with queries not present in the historic click log (unseen queries).

\section{Related work} \label{related}

This paper relates to the creation of effective dense retrievers methods. \textit{Dense Retrievers} (DRs)~\cite{xiong2020approximate,lin2020distilling,karpukhin2020dense,khattab2020colbert,hofstatter2021efficiently} utilise a BERT-based (or variation of) dual-encoder architecture to separately encode queries and passages into a shared embedding space. DRs have shown to have high effectiveness paired with low query latency across different retrieval tasks and datasets. However, current research directions in the DRs space focus on the development of effective training methods, especially related to negative sampling~\cite{xiong2020approximate,gao2021complement,lee2019latent}, and the integration of pseudo relevance feedback mechanisms~\cite{yu2021improving,li2021pseudo}. The exploitation in the context of DRs of implicit relevance signals such as clicks collected in a search engine click log has so far received no attention. Our paper aims to fill this gap and lay the foundations for this new line of research for DRs.

The development of effective techniques for the integration of PRF in DRs is relevant to our work from two angles. First, both that work and ours attempt to improve the query representation by exploiting signals in addition to those provided by the query alone, though the signal we use is different from that considered by PRF. Second, the way in which  we integrate the historic click signal into DRs is fundamentally similar, in its more basic instantiation, to the vector PRF methods for DRs devised by~\citeauthor{li2021pseudo}~\cite{li2021pseudo}. Specifically, they investigated two simple instantiations, Average and Rocchio, that perform PRF with DRs in a zero-shot manner, i.e. without requiring the training of a new neural model or fine tuning of an existing one. Alternative, more complex approaches for integrating PRF into DRs exist. For example, Yu et al.~\cite{yu2021improving} replaced the query encoder in ANCE~\cite{xiong2020approximate} by purposely training a new PRF query encoder. This new PRF query encoder concatenates the original query text and the text from the PRF passages to form a new query, which is then encoded into a new query vector and used to perform matching. However, a recent study conducted by~\citeauthor{li2022improving}~\cite{li2022improving} has shown that the same ANCE-PRF training regime may not necessarily apply to more advanced dense retrievers such as TCT-ColBERTv2~\cite{lin2020distilling} and DistilBERT KD TASB~\cite{hofstatter2021efficiently}.

A fundamental difference however exists between PRF methods for DRs and the methods we propose in this paper, aside from the type of relevance signal that they consider. PRF methods are commonly part of a two-stage pipeline: a first round of retrieval is performed using the original query; this is then used as relevance signal to perform a second round of retrieval. The use of this second stage adds to the overall query latency. Our methods instead are single-stage: they directly modify the query representation exploiting the click signals from previous queries and retrieve using such modified representation. Thus, the methods we propose here have minimal or no effect on the query latency of the underlying DRs.


In investigating how to exploit historic click signals for improving DRs, we make two key connections to the area of online and counterfactual LTR~\cite{zhuang2020counterfactual,oosterhuis2018differentiable,jagerman2019model,joachims2017unbiased,ovaisi2020correcting}. First, we rely on some of the components of the standard evaluation practice in that area, such as the use of click models~\cite{chuklin2015click,guo2009efficient} to simulate clicks and therefore generate the historic click log~\cite{joachims2017unbiased,joachims2002optimizing,jagerman2019model,oosterhuis2021robust}. Borrowing methods from this previous research allows us to have the foundational tools to address Challenge 1, outlined in section 1. Second, to address  the presence of bias in the click signal (Challenge 2 in section 1), we again tap into the research in counterfactual LTR, and integrate the notion of counterfactual de-biasing in the proposed Rocchio mechanism for DRs. 
Our work bears a resemblance of tabular models in LTR, where a ranker is learnt specifically for an individual query from
direct interactions with the user, i.e., they model each query as an independent ranking problem, but therefore cannot generalise beyond that query~\cite{zoghi2016click,oosterhuis2021robust}.
Despite these connections, we note that online/counterfactual LTR methods themselves are not comparable to  DRs (thus of course are not considered as baseline for evaluation). In fact, DRs may be part of a more complex pipeline that involve such LTR approaches, and perhaps they can even be used as one of the many features LTR methods exploit. 

One of the challenges we had to address in our work was the absence of a dataset adequate to train and evaluate DRs, like TREC DL, and that also contained historic click logs. We note that the MS MARCO dataset comes with an associated click log: ORCAS~\cite{craswell2020orcas}. However, this click log refers to the document part of MS MARCO, and thus it is difficult to derive from it clicks that are associated to the passages in the MS MARCO and TREC DL datasets that are typically used for researching DRs on the passage ranking task. We note that only a subset of the documents in MS MARCO has a mapping to the passage part of the dataset.
Another limitation of ORCAS is that clicks are recorded as query-document pairs and no information regarding the rank position the document was displayed at in the SERP has been recorded for that click. This affects our ability to derive position bias information. Hence we cannot use ORCAS for our experiments.

\section{Method} \label{method}



Next we introduce our counterfactual Rocchio (CoRocchio) method which exploits the click signal from  an historic query log to compute a new dense representation of the user query. CoRocchio relies on the framework of dense retrievers: queries and passages are encoded using a dense pre-trained deep language model encoder (e.g. ANCE, DPR, etc.) and retrieval is performed by computing the inner product between the dense representation of the query and that of the passages. In CoRocchio, however, the query representation is computed by aggregating the dense representation of the user's query with that of passages in the historic query log that have been clicked for that query (Figure~\ref{fig:implicit}).

CoRocchio relies on two key intuitions: (1) The passages that have received a click in the query log are often the user-preferred passages for that query; (2) Aggregating the dense representation of the query with these of the user-preferred passages will result in a new query representation that better matches user preference. 

The first intuition has been shown valid in previous work on learning to rank~\cite{joachims2002optimizing}, online learning to rank and evaluation~\cite{oosterhuis2018differentiable,zhuang2020counterfactual,oosterhuis2017sensitive,yue2009interactively} and counterfactual learning to rank~\cite{oosterhuis2021robust,joachims2017unbiased}.

The second intuition is also convincing: the correct aggregation of the original dense representation of the query with these of the user-preferred passages will move the query representation towards those of the preferred passages. Thus, other passages that have a similar dense representation will obtain a larger score with the new query representation than the old one -- and these passages are highly likely to also be preferred by the users.

\subsection{DRs with Implicit Feedback}


Relying on the above intuitions, we first define the Rocchio algorithm with DRs for exploiting historic clicks. This method is an adaptation of Li et al.'s Rocchio PRF for DRs~\cite{li2021pseudo} to the situation in which pseudo relevance feedback is replaced by the implicit feedback derived from the click-through signal.

Let $c(p_i)\in{\{0,1\}}$ be the binary function indicating if a user clicked on the passage $p_i$ ($c(p_i)=1$) or not ($c(p_i)=0$). Furthermore, let $\vec{q}$ and $\vec{p}$ be the query $q$ dense representation and passage $p$ dense representation, respectively. Each time the query is encountered in the query log, a ranking $r_q$ of passages has been generated by the baseline DR and clicks collected. The set of all historic rankings for query $q$ is indicated with $R_q$.

Now, let's consider the case in which query $q$ is issued to the online system and the historic query log wants to be used to improve the effectiveness of the baseline DR ranker. To do so, we use the Rocchio formula below to create a new query dense representation $\vec{q}^{'}$ from click information:



\begin{equation}
	\begin{split}
		\vec{q}^{'} & =\frac{1}{|R_q|} \sum_{r_q\in{R_q}}  \left[ \alpha \cdot \vec{q} + \beta \cdot \sum_{p_i\in{r_q}} \vec{p}_i \cdot c(p_i)  \right] \\
		& = \alpha \cdot \vec{q} + \frac{\beta }{|R_q|} \cdot \sum_{r_q\in R_q}\sum_{p_i\in r_q} \vec{p}_i \cdot c(p_i) \label{rocchio}
	\end{split}
\end{equation}

where $\alpha$ and $\beta$ are the Rocchio parameters: $\alpha$ controls the weight assigned to the original query and $\beta$ controls the weight assigned to the aggregated dense representation from the clicked passages. This new query representation $\vec{q}^{'}$ is then used to rank passages according to the inner product between $\vec{q}^{'}$ and the passages' dense representations.

In the edge case when all the relevant passages for a query have been displayed in the SERPs in the historic logs and users have clicked on all relevant passages, and have not clicked any not relevant ones, then new query representation $\vec{q}_{c}^{*}$ only aggregates relevant passages:

\begin{equation}\label{eq:perfect}
	\vec{q}^{*}= \alpha \cdot \vec{q} + \frac{\beta }{|R_q|} \cdot \sum_{r\in{R_q}}\sum_{p_i\in{r_q}} \vec{p}_i \cdot y(p_i) 
\end{equation}
where $y(p_i)=1$ if $p_i$ is relevant or $y(p_i)=0$ otherwise.

\subsection{CoRocchio: Unbiased Aggregation} \label{corocchio_proof}
Equation~\ref{rocchio} treats every click equally. However, the click signal presents a number of biases, the strongest being position bias~\cite{yue2010beyond,joachims2017unbiased}. Position bias manifests as the tendency of the users to click more often (or more likely) on passages that are ranked earlier in the SERP. 
Not accounting for position bias, as it is done in Equation~\ref{rocchio}, may lead to suboptimal effectiveness. 

Next, we mathematically prove that the Rocchio method for DRs of Equation~\ref{rocchio} is biased with respect to position and thus leads to suboptimal results. Following previous work in counterfactual learning to rank~\cite{oosterhuis2021robust,joachims2017unbiased}, we develop this demonstration in the restrictive setting that clicks occur only and only if a passage is relevant and examined by the user. We do this ideal and unrealistic assumption for convenience only in this mathematical proof; we note this assumption is well accepted by the counterfactual learning to rank community. Extensions of this proof to nosier and more realistic conditions are possible (though not trivial) but this aspect is out of scope for this paper.

Let assume that the probability of a passage $p_i$ to be examined by a user (known as propensity) only depends on its rank position in the SERP and that the user will click on every relevant passage they examined (i.e., no click noise). Thus, the probability of a click on $d_i$ is:


\begin{equation}
	P\left(c(p_i)\right) = P(o_i) \cdot y(p_i)
\end{equation}
where $P(o_i)$ is the examination propensity of rank position $i$. If we treat clicks as an unbiased relevance signal, then the expectation of $\vec{q}$ is:
\begin{equation}
			\begin{split}
	\mathbb{E}_{o}\left[  \vec{q}^{'} \right] & = \mathbb{E}_{o}\left[  \alpha \cdot \vec{q} + \frac{\beta }{|R_q|} \cdot \sum_{r_q\in R_q}\sum_{p_i\in{r_q}} \vec{p}_i \cdot c(p_i) \right] \\
	& = \alpha \cdot \vec{q} + \frac{\beta }{|R_q|} \cdot \sum_{r_q\in R_q}\sum_{p_i\in r_q} \vec{p}_i \cdot P(o_i) \cdot y(p_i) \neq \vec{q}^{*}
		\end{split}
\end{equation}
This means $\vec{q}^{'}$ is a biased estimate of $\vec{q}^*$. According to empirically collected data~\cite{guan2007eye}, higher rank positions usually have higher examination probability. Thus, $\vec{q}^{'}$ biasedly assigns higher weights to clicked passages displayed at earlier ranks in the SERP.

To overcome the effect of position bias in the click log, inspired by counterfactual learning to rank practice, we propose the Counterfactual Rocchio (CoRocchio) which uses the inverse propensity scoring (IPS)~\cite{joachims2017unbiased,horvitz1952generalization} to de-bias the click signal:
\begin{equation}
\mbox{CoRocchio}\left( \vec{q}, P(o) \right) = \alpha \cdot \vec{q} + \frac{\beta }{|R_q|} \cdot \sum_{r_q\in R_q}\sum_{p_i\in r_q } \frac{\vec{p}_i}{P(o_i)} \cdot c(p_i) 
\end{equation}
CoRocchio generates an unbaised estimate of $\vec{q}^{*}$ if every relevant passage ($y(p_i)=1$) has a positive examination propensity ($P(o_i) > 0$):

	\begin{align*}
	&\mathbb{E}_{o}\left[ \mbox{CoRocchio}\left(\vec{q}^{'}, P(o) \right) \right] \\
	&= \mathbb{E}_{o}\left[ \alpha \cdot \vec{q} + \frac{\beta }{|R_q|} \cdot \sum_{r_q\in R_q}\sum_{p_i\in r_q} \frac{\vec{p}_i}{P(o_i)} \cdot c(p_i)  \right] \\
	&= \alpha \cdot \vec{q} + \frac{\beta }{|R_q|} \cdot \sum_{r_q\in R_q}\sum_{p_i\in r_q} \frac{\vec{p}_i \cdot P(o_i) }{P(o_i)} \cdot y(p_i) \\
	&= \alpha \cdot \vec{q} + \frac{\beta }{|R_q|} \cdot \sum_{r_q\in R_q}\sum_{p_i\in r_q} \vec{p}_i \cdot y(p_i)  = \vec{q}^{*} \numberthis \label{eqn}\\
	\end{align*}

Intuitively, CoRocchio assigns more weights to clicked passages that appear at lower ranks in the SERP by dividing their examination propensity, hence discounting the position bias. This requires to know the propensity $P(o)$ a priori. Following standard counterfactual learning to rank experimental settings~\cite{joachims2017unbiased} and for simplicity, we assume in our experiments the propensities are known and regard propensity estimation as being beyond the scope of this paper. However, we note propensity estimation is a well-studied area on its own and many recent studies have considered estimating such propensities from historical click-logs~\cite{agarwal2017effective,fang2019intervention,agarwal2019estimating}.

\subsection{Dealing with Unseen Queries in CoRocchio}\label{sec:dealing_unseen}
Above, we have shown CoRocchio can unbiasedly estimate the optimal query representation for dense retriever from an historic click log. One drawback of CoRocchio is that it can only be used for queries that appear in the historic click log, resembling a tabular-based ranker~\cite{zoghi2016click,oosterhuis2021robust} (at the next of not being LTR-based): it cannot deal with unseen queries. 
This is acceptable for popular queries: they would be logged many times by the search engine. However, the majority of queries logged by search engines are tail queries or have never been logged before~\cite{silverstein1999analysis}.

 In order to generalise CoRocchio to unseen queries, we devise the CoRocchio with Approximate Nearest Neighbor Query (CoRocchio-ANN) algorithm. CoRocchio-ANN is built on the assumption that similar queries would have similar dense representation encodings given by the DR query encoder.
 More specifically, we use the baseline DR encoder to encode the unseen query into its dense representation $\vec{q_u}$. Then we use $\vec{q_u}$ to search for the top-k nearest neighbor queries in the historic query log. Let $Q$ be the query set of top-k nearest neighbor ($|Q| = k$); then the new query vector for $\vec{q_u}$ is:
 
	\begin{align*}
	&\mbox{CoRocchio-ANN}(\vec{q_u}, P(o))\\
	&= \alpha \cdot \vec{q_u} + \frac{\beta }{|Q| \cdot |R_q|} \cdot \sum_{q\in{Q}}\sum_{r_q\in R_q}\sum_{p_i\in r_q} \frac{\vec{p}_i}{P(o_i)} \cdot c(p_i) \numberthis \label{eqn}\\
	\end{align*}
In other words, we use the average unbiased passage representation aggregation for the top-k nearest neighbor queries to compute the new query representation for the unseen query. 

We note that although all the methods proposed in this section only considered clicked passages in the historic query log as relevant implicit feedback, the original Rocchio algorithm formulation allows for negative feedback as well. We do not explicitly integrate negative implicit feedback, i.e. unclicked passages, into our CoRocchio algorithms because it is not trivial to de-bias the user's behaviour on passages that are not clicked. This is because it is not straightforward to ascertain why a user has not clicked on a passage: is it because the passage is not relevant or because the user did not examine the passage? Nevertheless, we realize that there have been recent attempts to de-bias unclick signals that have used IPS-like algorithms~\cite{wang2021non,zhuang2022reinforcement}: these can potentially be integrated into our CoRocchio algorithm to further leverage the user's negative implicit feedback for dense retrievers. We leave this to future exploration.
\section{Experimental settings} \label{experiments}
To study the impact implicit feedback in terms of click signal has on the effectiveness of DRs and the effectiveness of our proposed CoRocchio, we devise a number of empirical experiments aimed at investigating: (1) the effectiveness of CoRocchio when using the signal from the historic click log (biased and unbiased) for queries present in the log (results in section~\ref{res-corocchio}), and (2) the effectiveness of CoRocchio-ANN when exploiting the historic click signal to answer previously unseen queries (results in section~\ref{res-corocchio-ann})

%

\subsection{Dataset}\label{sec:dataset}
We use the TREC Deep Learning Track Passage Retrieval Task 2019~\cite{craswell2019overview} (DL 2019) and 2020~\cite{craswell2020overview} (DL 2020). DL2019 and DL2020 contain 43 and 54 judged queries each. The relevance judgements for both datasets range from 0 (not relevant) to 3 (highly relevant); however, relevance label 1 indicates passages on-topic but not relevant and hence we relabel these passages to 0 when we compute binary relevance metrics (e.g., MAP, Recall).
 The passage collection is the same as the MS MARCO passage ranking dataset~\cite{nguyen2016ms}, a benchmark English dataset for ad-hoc retrieval tasks with around 8.8 million passages. The difference between TREC DL
and MS MARCO is that queries in TREC DL have several judgements per query (215.3/210.9 on average for 2019/2020), instead of only an average of one judgement per query for MS MARCO.

We do not use MS MARCO dataset to evaluate methods in this work. This is because the number of judgments per query is crucial in our experiments since we follow the standard unbiased LTR practice, which relies on relevance judgments and a click model (more details in section~\ref{sec:synthetic_click}) to synthetically generate user clicks. Since MS MARCO contains only one judgment per query, if used in this context the click model will generate extremely biased click data, largely differing from real-world user behaviour. 

\subsection{Synthetic User Behavior}\label{sec:synthetic_click}
To fully control users' biases and noise so that algorithms can be tested under different, controllable conditions, it is common for research in online LTR and counterfactual LTR to simulate users’ clicks by relying on the relevance labels recorded in offline datasets~\cite{zhuang2020counterfactual,oosterhuis2018differentiable,jagerman2019model,joachims2017unbiased,ovaisi2020correcting,zhuang2022reinforcement}. Following this practice, we use a click model to simulate click behaviour and create a synthetic click log.

In our experiments, clicks are simulated based on two fixed variables: the click probability and the position bias. The click probability $P(c(p=1)|o=1, y(p))$ is the probability of a user clicking on a passage after exam it. This probability is conditioned on the passage's relevance label $y(p)$. We set two types of click behaviour: \textit{perfect} and \textit{noisy}. The click probability of the \textit{perfect} click behaviour is proportional to the relevance level of the passages, and has $0$ probability for non-relevant passages. This simulates an ideal user that is able to always determine the relevance of a passage in the SERP. The \textit{noisy} click behaviour mimics instead a realist behaviour on SERPs by assigning a small click probability to non-relevant passages and a small skip probability to relevant passages. Table~\ref{click_model} provides the click probabilities for the two user models.

Position bias is modelled by the passage observation probabilities $P(o_{i}=1) $; we assume the observation probabilities only depend on the rank position of the passage and set these probabilities as:
\begin{equation}\label{eq:propensity}
	P(o_{i}=1) = \left( \frac{1}{i} \right)^\eta
\end{equation}
where $i$ is the rank position of the given passage $d$ and $\eta$ is a parameter that determines the level of position bias. Following Joachims et al.~\cite{joachims2017unbiased} and Jagerman et al.~\cite{jagerman2019model}, we set $\eta=1$ for our main experiments, while we investigate the impact of different $\eta$ values in Section~\ref{res-corocchio}.
Thus, the probability of a click occurring on a passage at rank $i$ in the result list is:
\begin{equation}
	P(c(p)=1) = P(c(p)=1|o_i=1, y(p)) \cdot P(o_i=1)
\end{equation}

With the click model, we then create an historic click log by issuing each query in the TREC DL datasets to the DRs and record the top 10 retrieved passages and the simulated clicks on these passages. We repeat this simulation 1,000 times for each query to log enough click signal; although we note that a lower or higher number of simulations provide similar trends and observations to the ones reported here.

\begin{table}
	\centering
	\caption[centre]{Click probabilities for different user behaviours.}\label{click_model} 
	\begin{tabular}{rcccc}
		\toprule
		&  \multicolumn{4}{c}{$P(c(p)=1|o=1, y(p))$} \\
		\midrule
		$y(p)$ & 0& 1 & 2 & 3 \\
		\midrule
		\textit{perfect} & 0.0 & 0.0 & 1.0 & 1.0\\
		\textit{noisy} & 0.2 & 0.4 & 0.8 & 0.9 \\
		\bottomrule
	\end{tabular}
\end{table}

\subsection{Dataset Augmentation with Synthetic Query Generation}\label{sec:synthetic_data}
To evaluate CoRocchio-ANN, which allows to use DRs with implicit feedback in presence of queries not seen in the historic query log, a dataset with unseen queries that are related to those that have been observed in the log is required; and these queries need also to have relevance judgements. Because TREC DL has only a small set of judged queries, and they are unrelated to each other, withholding  a subset of TREC DL queries is as unseen queries is not possible. We then have to devise a different avenue to generate a dataset that allows to study this aspect of CoRocchio. To this aim, we adapt the docTquery-T5 method~\cite{nogueira2019doc2query} to augment the current TREC DL datasets with unseen, but related queries with associated relevance judgements.

The docTquery-T5 method is a T5 language model~\cite{raffel2020exploring} that is fine-tuned on the task of generating relevant queries for any given passage. It has been shown effective at generating high quality queries for the task of corpus summarisation ~\cite{surita2020can} and passage expansion~\cite{mallia2021learning,zhuang2021fast,lin2021few}. We use  docTquery-T5 to generate a query from each passage that has been judged relevant~\footnote{We treat passages with relevance levels 2 and 3 as relevant passages and exclude those with label 1.} for one of the original queries in the TREC DL dataset. We assume that the passages relevant to the original query for which a passage was taken for query generation, are also relevant to the generated query and we share the same relevant passage set between the original query and its associated generated queries. With docTquery-T5, it may be possible that the generated query and the original query are identical: in this case we sample again a query from docTquery-T5 to ensure the query generated from a passage is different from the original query for that passage.  Since we generate a query for each relevant passage given an original TREC DL query, on average we generate 58 synthetic queries per each original query in TREC DL 2019 and 31 for TREC DL 2020. In total, we generated 2,501 extra queries for DL2019 and 1,666 extra queries for DL2020.  In our experiments, we split the augmented query set into a seen query set (80\% of the generated queryes) and an unseen query set (20\%). The seen query set is used to simulate the historic click log while the unseen query set is used for evaluating CoRocchio-ANN.

Table~\ref{table:case} reports examples of original and generated queries. As expected, the generated queries are similar to the original query. While there are cases in which the original and generated queries differ enough that some of the documents relevant to the original query may not be relevant to the generated query, we empirically observed these to not be frequent enough to be a main source of error in the evaluation -- we note that even in a high quality collection such as TREC DL relevance assessments are at time noisy~\cite{arabzadeh2021shallow}.

\begin{table}[t]
	\caption {Examples of original queries in DL2019 (the first two rows) and DL2020 (the second two rows) vs the queries generated from one of their relevant passages by docTquery-T5.}
	\label{table:case}
	\begin{small}
		\begin{tabular}{  p{3cm} | p{4.8cm}  }
			\toprule
			\bf Original queries& \bf  Generated queries  \\
			\hline
			do goldfish grow & how big do shubunkin fish get    \\
			\hline
			what is wifi vs bluetooth & what is the difference between bluetooth and wifi?    \\
			\hline
			who is aziz hashim & who is hashim franchisor    \\
			\hline
			do google docs auto save & how do you save google docs updates \\
			\bottomrule
		\end{tabular}
	\end{small}
\end{table}

\begin{table*}[t]
	\caption {Logged query results. $^{\dag}$ indicates statistically significant differences (p=0.05 ) between DRs with click signal vs. their respective DRs with PRF signal. Differences between Rocchio vs. CoRocchio are marked with $^\star $ if statistically significant.} \label{seen-results} 
	\centering
		\resizebox{\textwidth}{!}{
		\begin{tabular}{ l |llll|llll }
			\toprule
			& \multicolumn{4}{c|}{\bf TREC DL2019} & \multicolumn{4}{c}{\bf TREC DL2020} \\ \toprule
			\bf Method   & \bf nDCG@10 & \bf nDCG@100  & \bf Recall@1000  &  \bf MAP         & \bf nDCG@10 & \bf nDCG@100  &\bf  Recall@1000  &  \bf MAP       \\ \midrule
			BM25   &0.4973&0.4981&0.7450&0.2901&0.4876&0.4914&0.8031&0.2876\\
			ANCE~\cite{xiong2020approximate} &0.6452&0.5540&0.7554&0.3710&0.6458&0.5679&0.7764&0.4076\\
			TCT-ColBERTv2~\cite{lin2021batch} &0.7204&0.6318&0.8261&0.4469&0.6882&0.6206&0.8429&0.4754\\
			\midrule
			\textbf{Pseudo-relevance feedback}      & && &&  &&   &   \\
			BM25 + RM3~\cite{abdul2004umass} &0.5231&0.5263&0.7792&0.3377&0.4808&0.5145&0.8286&0.3056\\
			ANCE + VPRF~\cite{li2021pseudo}
			&0.6561&0.6018&0.7562&0.4278&0.6221&0.5636&0.7875&0.4098\\
			ANCE-PRF~\cite{yu2021improving} &0.6807	&0.5950	&0.7912	&0.4253	&0.6948	&0.5947	&0.8148	&0.4452\\
			TCT-ColBERTv2 + VPRF~\cite{li2021pseudo} &0.6982	&0.6556	&0.8633	&0.4797	&0.6678	&0.6058	&0.8514	&0.4697\\
			
			\midrule
			\textbf{Exploits Implicit feedback}  & && && && &   \\
			\textbf{Perfect and unbiased user}   & && && && &   \\
			ANCE + Rocchio &0.7543$^{\dag}$&0.6557$^{\dag}$&0.8144&0.5045$^{\dag}$&0.7438$^{\dag}$&0.6227&0.8170&0.5077\\
			TCT-ColBERTv2 + Rocchio &0.7963$^{\dag}$&0.7233&0.9001&0.5778$^{\dag}$&0.8095$^{\dag}$&0.6998$^{\dag}$&0.9033$^{\dag}$&0.5935$^{\dag}$\\
			\textbf{Perfect and biased user}   & && && && &   \\
			ANCE + Rocchio &0.7243&0.6393$^{\dag}$&0.8047&0.4764&0.7199&0.6140&0.8164&0.4848\\
			ANCE + CoRocchio &0.7522$^{\dag}$$^\star$&0.6558$^{\dag}$$^\star$&0.8149$^\star$&0.5033$^{\dag}$$^\star$&0.7413$^{\dag}$$^\star$&0.6227&0.8175&0.5080\\
			TCT-ColBERTv2 + Rocchio &0.7883$^{\dag}$&0.7121$^{\dag}$&0.8948&0.5592$^{\dag}$&0.7893$^{\dag}$&0.6846$^{\dag}$&0.8970&0.5571$^{\dag}$\\
			TCT-ColBERTv2 + CoRocchio &0.7963$^{\dag}$&0.7237$^{\dag}$&0.9004&0.5774$^{\dag}$&0.8093$^{\dag}$$^\star$&0.6996$^{\dag}$$^\star$&0.9014$^{\dag}$&0.5935$^{\dag}$\\
			
			\textbf{Noise and unbiased user}   & && && && &   \\
			ANCE + Rocchio &0.7167&0.6314$^{\dag}$&0.7848&0.4665&0.6981&0.6057&0.8070&0.4698\\
			TCT-ColBERTv2 + Rocchio &0.7759$^{\dag}$&	0.7062$^{\dag}$&	0.8813&	0.5495$^{\dag}$&	0.7559$^{\dag}$&	0.6647$^{\dag}$&	0.8767$^{\dag}$&	0.5424$^{\dag}$\\
			\textbf{Noise and biased user}   & && && && &   \\
			ANCE + Rocchio &0.7019&	0.6218&	0.7921&	0.4589&	0.6877&	0.6044&	0.8114&	0.4593\\
			ANCE + CoRocchio &0.7119&	0.6295$^{\dag}$&	0.7836&	0.4657&	0.6965&	0.6049&	0.8060&	0.4688\\
			TCT-ColBERTv2 + Rocchio &0.7688$^{\dag}$&	0.7028$^{\dag}$&	0.8830&	0.5381&	0.7444$^{\dag}$&	0.6611$^{\dag}$&	0.8828$^{\dag}$&	0.5276$^{\dag}$\\
			TCT-ColBERTv2 + CoRocchio &0.7730$^{\dag}$&	0.7054$^{\dag}$&	0.8816&	0.5503$^{\dag}$&	0.7568$^{\dag}$&	0.6662$^{\dag}$&	0.8827$^{\dag}$&	0.5446$^{\dag}$\\
			\bottomrule
		\end{tabular}
	}
	\label{table:seen_results}
\end{table*}

\subsection{Baselines and Metrics}
We are the first to study how to exploit implicit feedback, in particular the click-through signal, with DRs -- thus baselines are limited.

We study the use of the proposed CoRocchio method and its variants Rocchio (without counterfactual de-biasing) and CoRocchio-ANN (for unseen queries) in the context of two DRs: ANCE~\cite{xiong2020approximate} and TCT-ColBERTv2~\cite{lin2021batch}, although the method could be generalised to other DRs. ANCE and TCT-ColBERTv2 then naturally represent two baselines for CoRocchio. We also report the effectiveness of  BM25 + RM3, a representative bag-of-words PRF method, since it is often used as a strong bag-of-word baseline in DRs research.

Strictly speaking CoRocchio is not a relevance feedback mechanism of the like of PRF because, at the net of replacing the pseudo relevance signal with the implicit relevance signal, CoRocchio does not consider the relevance feedback from an initial round of retrieval , but the implicit feedback from an historic click log (and thus past searches, possibly performed by other users). Despite this, the CoRocchio aggregation function is similar to that of the vector PRF (VPRF) methods proposed for DRs~\cite{li2021pseudo}, especially if no counterfactual de-biasing is used. We therefore also use VPRF as a baseline: this method can be applied to any DR. This allows us to establish the value of the implicit relevance signal over the PRF signal in the context of DRs, since the scoring method results to be similar (at the net of the counterfactual de-bias). We also then report as baseline a different way of performing PRF with DRs: the recently proposed ANCE-PRF~\cite{yu2021improving}. Instead of aggregating the dense representations of the feedback passages as in CoRocchio and VPRF, ANCE-PRF trains a new encoder to be able to generate effective representations of the concatenation of the original text of the query and that of the feedback passages. This method is only available for the ANCE DR.

For all Rocchio-based methods (our Rocchio, CoRocchio, CoRocchio-ANN and VPRF), we set $\alpha=0.4$ and $\beta=0.6$ for fair comparison. Our method and all baselines are implemented using Pyserini~\cite{lin2021pyserini} and the code will be made publicly available upon acceptance. We compare the methods with respect to nDCG@10, nDCG@1000, Recall@1000 and MAP, which are commonly used in TREC DL. Statistical differences between methods' results are computed using two-tailed paired t-test with Bonferroni correction.

\captionsetup[sub]{font=small,labelfont={bf,sf}}
\begin{figure*}
	\begin{subfigure}{.45\textwidth}
		\centering
		\includegraphics[width=\linewidth]{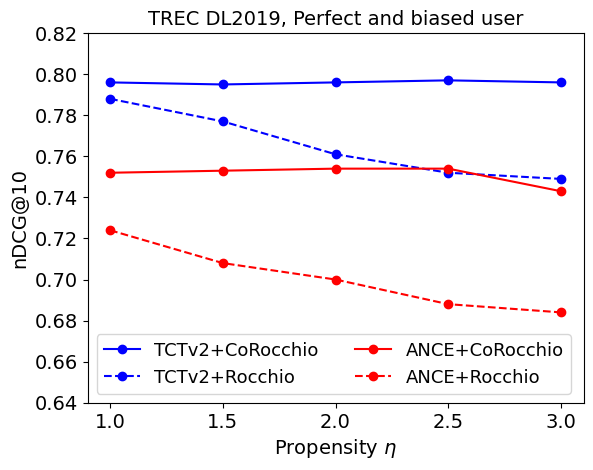}
		\caption{DL2019, perfect and biased user.}
		\label{fig:sfig1}
	\end{subfigure}%
	\begin{subfigure}{.45\textwidth}
		\centering
		\includegraphics[width=\linewidth]{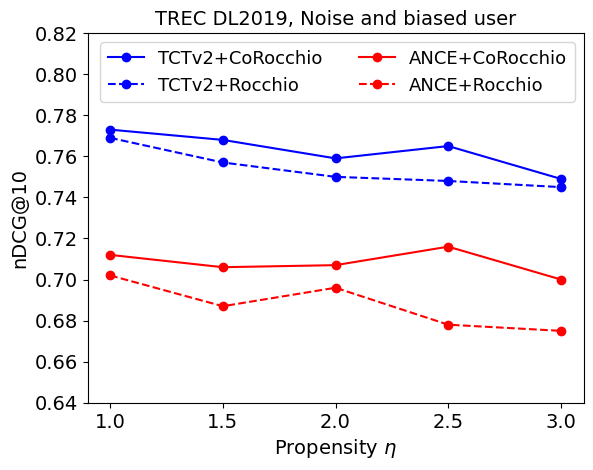}
		\caption{DL2019, noise and biased user.}
		\label{fig:sfig2}
	\end{subfigure}
	\begin{subfigure}{.45\textwidth}
		\centering
		\includegraphics[width=\linewidth]{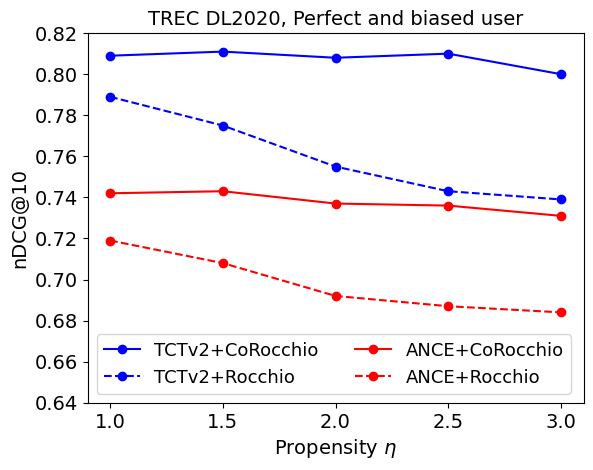}
		\caption{DL2020, perfect and biased user.}
		\label{fig:sfig3}
	\end{subfigure}
	\begin{subfigure}{.45\textwidth}
		\centering
		\includegraphics[width=\linewidth]{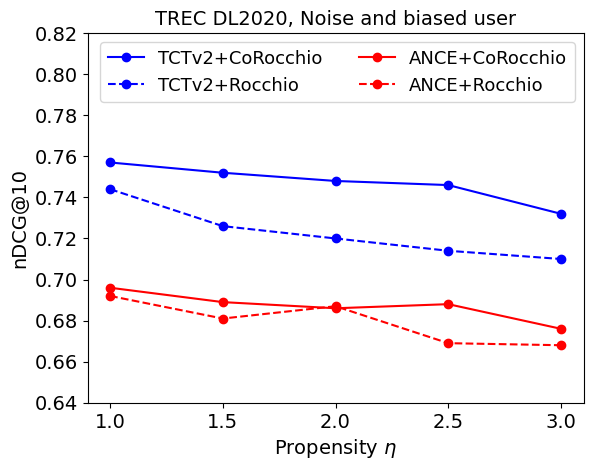}
		\caption{DL2020, noise and biased user.}
		\label{fig:sfig4}
	\end{subfigure}
	\caption{nDCG@10 score of ANCE and TCT-ColBERTv2 with Rocchio or CoRocchio under different propensity settings.}
	\label{fig:fig}
\end{figure*}

\section{Results} \label{results}

\subsection{Implicit Feedback and CoRocchio} \label{res-corocchio}
We first consider the results obtained when using the proposed method for DRs to exploit the historical click log for queries observed in the log itself (logged queries).  
The main results for such logged queries are presented in Table~\ref{seen-results}. 

Let us briefly consider the baselines. Comparison between BM25, BM25+RM3, ANCE and TCT-ColBERTv2 reinforce previous findings in the literature: DRs are generally more effective than baseline bag-of-words-methods (including with PRF). Similarly, the use of PRF for DRs generally improves DR baselines for Recall@1000 and MAP; however, for the shallow metrics such as nDCG@10, the improvement of PRF is unstable~\cite{li2021pseudo}. For example, ANCE + VPRF and TCT-ColBERTv2 + VPRF have lower nDCG@10 than the respective DRs without PRF; the only PRF method that consistently outperforms the corresponding DR baseline is ANCE-PRF~\cite{yu2021improving}. We note that VPRF is a zero-shot method (untrained) while ANCE-PRF requires training, so these results are unsurprising.


Next, let us consider the results obtained when exploiting the click logs. 

When the click model is that of the perfect user, and no bias is injected (\textbf{Perfect and unbiased user}), then the users' clicks recorded in the log have been on all and only the relevant passages present in the SERPs. This means there is no click noise and no position bias ($P(o_i)=1$ always) and thus CoRocchio reduces to Rocchio (thus we do not report CoRocchio in Table~\ref{seen-results} for this setting). Under this setting, exploiting the historic click signal in DRs using our Rocchio returns significant effectiveness boosts across all datasets and metrics for both of the considered DRs.
This result is to be expected: in this setting the click signal is a very clear indicator of the relevant passages, despite this is only collected only for a subset of the relevant passages for a query, i.e. those displayed in the SERP (10 passages) displayed in the historic click log, while the methods are evaluated across the full ranking (top 1,000 passages). However, we still believe this edge case result is valuable as it suggests that, for IR applications where the user real preference feedback is available, such as during the screening process in technology assisted reviews~\cite{oard2013information,mcdonald2018active,zhang2020evaluating}, DRs can be adapted to exploit such a signal in a very effective manner.

 \begin{table*}[t]
	\caption {Unseen query results. $^{\dag}$ indicates statistically significant differences (p=0.05 ) between DRs with click signal vs. their respective DRs with PRF signal. Differences between Rocchio vs. CoRocchio are marked with $^\star $ if statistically significant.} \label{unseen-results}
	\centering
		\resizebox{\textwidth}{!}{
			\begin{tabular}{ l |llll|llll }
				\toprule
				& \multicolumn{4}{c|}{\bf TREC DL2019} & \multicolumn{4}{c}{\bf TREC DL2020} \\ \toprule
				\bf Method   & \bf nDCG@10 & \bf nDCG@100  & \bf Recall@1000  &  \bf MAP         & \bf nDCG@10 & \bf nDCG@100  &\bf  Recall@1000  &  \bf MAP       \\ \midrule
				BM25   &0.2996	&0.2578	&0.4258	&0.1409	&0.2164	&0.2128	&0.4557	&0.1140\\
				ANCE~\cite{xiong2020approximate} &0.4024	&0.3400	&0.4622	&0.2156	&0.3251	&0.2725	&0.4802	&0.1679\\
				TCT-ColBERTv2~\cite{lin2021batch} &0.4203	&0.3690	&0.5211	&0.2417	&0.3298	&0.2888	&0.5336	&0.1848\\
				\midrule
				\textbf{Pseudo-relevance feedback}      & && &&  &&   &   \\
				BM25 + RM3~\cite{abdul2004umass} &0.3302	&0.3031	&0.4971	&0.1850	&0.2430	&0.2474	&0.5124	&0.1443\\
				ANCE + VPRF~\cite{li2021pseudo} &0.4323	&0.3840	&0.5084	&0.2575	&0.3445	&0.3071	&0.5250	&0.2020\\
				ANCE-PRF~\cite{yu2021improving} & 0.4297	&0.3710	&0.5008	&0.2440	&0.3482	&0.3006	&0.5201	&0.2009 \\
				TCT-ColBERTv2 + VPRF~\cite{li2021pseudo} &0.4523	&0.4142	&0.5769	&0.2906	&0.3411	&0.3284	&0.5801	&0.2267\\
				\midrule
				\textbf{Exploits Implicit feedback}  & && && && &   \\
				\textbf{Perfect and unbiased user}   & && && && &   \\
				ANCE + Rocchio-ANN &0.6404$^{\dag}$	&0.5155	$^{\dag}$&0.6036$^{\dag}$	&0.3625$^{\dag}$	&0.5999$^{\dag}$	&0.4569$^{\dag}$	&0.6489$^{\dag}$	&0.3292$^{\dag}$\\
				TCT-ColBERTv2 + Rocchio-ANN &0.6904$^{\dag}$	&0.5588$^{\dag}$	&0.6752$^{\dag}$	&0.4059$^{\dag}$	&0.6314$^{\dag}$	&0.5091$^{\dag}$	&0.7289$^{\dag}$	&0.3918$^{\dag}$\\
				\textbf{Perfect and biased user}   & && && && &   \\
				ANCE + Rocchio-ANN  &0.5872$^{\dag}$	&0.4768$^{\dag}$	&0.5767$^{\dag}$	&0.3284$^{\dag}$	&0.5287$^{\dag}$	&0.4120$^{\dag}$	&0.6168$^{\dag}$	&0.2884$^{\dag}$\\
				ANCE + CoRocchio-ANN  &0.6395$^{\dag}$$^\star$	&0.5152$^{\dag}$$^\star$	&0.6036$^{\dag}$$^\star$	&0.3625$^{\dag}$$^\star$	&0.5986$^{\dag}$$^\star$	&0.4564$^{\dag}$$^\star$	&0.6488$^{\dag}$$^\star$	&0.3288$^{\dag}$$^\star$\\
				TCT-ColBERTv2 + Rocchio-ANN  &0.6325$^{\dag}$	&0.5242$^{\dag}$	&0.6490$^{\dag}$	&0.3747$^{\dag}$	&0.5698$^{\dag}$	&0.4645$^{\dag}$	&0.6992$^{\dag}$	&0.3451$^{\dag}$\\
				TCT-ColBERTv2 + CoRocchio-ANN  &0.6897$^{\dag}$$^\star $	&0.5585$^{\dag}$$^\star $	&0.6757$^{\dag}$$^\star $	&0.4061$^{\dag}$$^\star $	&0.6309$^{\dag}$$^\star $	&0.5085$^{\dag}$$^\star $	&0.7288$^{\dag}$$^\star$	&0.3915$^{\dag}$$^\star $\\
				\textbf{Noise and unbiased user}   & && && && &   \\
				ANCE + Rocchio-ANN  &0.5935$^{\dag}$	&0.4919$^{\dag}$	&0.5992$^{\dag}$	&0.3402$^{\dag}$	&0.5342$^{\dag}$	&0.4273$^{\dag}$	&0.6349$^{\dag}$	&0.2941$^{\dag}$\\
				TCT-ColBERTv2 + Rocchio-ANN  &0.6323$^{\dag}$	&0.5324$^{\dag}$	&0.6686$^{\dag}$	&0.3815$^{\dag}$	&0.5635$^{\dag}$	&0.4801$^{\dag}$	&0.7218$^{\dag}$	&0.3517$^{\dag}$\\
				\textbf{Noise and biased user}   & && && && &   \\
				ANCE + Rocchio-ANN  &0.5635$^{\dag}$	&0.4671$^{\dag}$	&0.5788$^{\dag}$	&0.3174$^{\dag}$	&0.4996$^{\dag}$	&0.4028$^{\dag}$	&0.6157$^{\dag}$	&0.2730$^{\dag}$\\
				ANCE + CoRocchio-ANN  &0.5923$^{\dag}$$^\star$	&0.4919$^{\dag}$$^\star$	&0.5979$^{\dag}$$^\star$	&0.3403$^{\dag}$$^\star$	&0.5337$^{\dag}$$^\star$	&0.4274$^{\dag}$$^\star$	&0.6350$^{\dag}$$^\star$	&0.2941$^{\dag}$$^\star$\\
				TCT-ColBERTv2 + Rocchio-ANN  &0.6068$^{\dag}$	&0.5142$^{\dag}$	&0.6525$^{\dag}$	&0.3633$^{\dag}$	&0.5271$^{\dag}$	&0.4499$^{\dag}$	&0.7005$^{\dag}$	&0.3219$^{\dag}$\\
				TCT-ColBERTv2 + CoRocchio-ANN  &0.6327$^{\dag}$$^\star$	&0.5325$^{\dag}$$^\star$	&0.6687$^{\dag}$$^\star$	&0.3813$^{\dag}$$^\star$	&0.5630$^{\dag}$$^\star$	&0.4802$^{\dag}$$^\star$	&0.7224$^{\dag}$$^\star$	&0.3515$^{\dag}$$^\star$\\
				\bottomrule
			\end{tabular}
		}
	\label{table:unseen_results}
\end{table*}

When bias is added, even in the absence of click noise (\textbf{Perfect and biased user}), we observe decreases in the effectiveness of Rocchio, regardless of the DR. For example, nDCG@10 on DL2019 for ANCE + Rocchio goes from 0.7543 for the perfect and unbiased user to 0.7243 when bias is introduced; TCT-ColBERTv2 + Rocchio also experiences drops, and drops are indeed observed across all evaluation metrics and datasets. These results indicate that position bias has a negative impact on learning an effective dense query representation by exploiting the click log signal with our Rocchio method. On the other hand, when CoRocchio is used to obtain the new dense query representation that exploits the information in the query log, higher effectiveness is observed. In fact, after de-biasing and aggregating the click signal with CoRocchio, all DRs obtain results that are similar to those obtain in absence of click bias. These results empirically prove what we mathematically demonstrated in section~\ref{corocchio_proof}: that CoRocchio can remove the position bias present in the historic click log, delivering increased effectiveness for the employed DRs. 

 
 Next we investigate the impact of click noise, by considering the noisy click model in place of the perfect.  When no click bias is present (\textbf{Noise and unbiased user}), Rocchio and CoRocchio result to be identical (as when the setting Perfect and unbiased user was considered).  Comparing the results obtained in this setting with those for the perfect user in the unbiased case, we observe significant drops in effectiveness -- in fact these drops are larger than those imposed by the introduction of bias in the perfect click model. This suggests that, at least in the settings investigated in our simulations, click noise is more detrimental to DRs effectiveness than click position bias.
 
  This suggests that click noise has bigger negative impact for DRs that exploit historic click log.

  On the other hand, when both click noise and position bias are present in the click log (\textbf{Noise and biased user}) -- the most realistic amongst the settings considered --  we observe Rocchio for DRs obtains the lowest effectiveness compared to other settings (but still higher than baselines) across all metrics except  Recall@1000. The higher  Recall@1000 observed for this setting may be because the noisy clicks that occur at lower-ranked positions have a higher negative contribution to Recall@1000 than the positive contribution coming from correct clicks. When position bias is present, lower-ranked passages have a lower observation probability and thus attract fewer noisy clicks. 
  However, if CoRocchio is used in the Noise and biased user setting, we observe that it can effectively reduce, if not eliminate, the position  bias, aligning its result in this context to those obtained for noisy clicks and no bias. 
  Interestingly, also in this case we observe higher Recall@1000 than when noisy but unbiased clicks were used. 
  
  	\textit{\textbf{Influence of user propensity ($\eta$).}} In the previous experiments, to simulate user position bias on SERPs $\eta$ was set to 1. In Figure~\ref{fig:fig}, we report the influence of higher values of user position bias $\eta$ on the effectiveness of Rocchio and CoRocchio. For this, we use different values of $\eta$ to simulate different user observation probabilities, where higher $\eta$ means the users are more biased towards the top ranked passages.
  	When $\eta=3$, users will have less than $5\%$ of chances to observe passages beyond  rank 3 (extreme bias). For this experiment, we report the nDCG@10 achieved by different methods on both TREC DL2019 and 2020.
  	From Figure~\ref{fig:fig}, it is clear that our CoRocchio outperforms Rocchio across all  datasets, models and $\eta$ values, empirically demonstrating the benefits of de-biasing  position bias. 
  	We also note that, under the perfect and biased user setting, the effectiveness of Rocchio decreases with the increase of $\eta$.
  	This shows the negative impact of user position bias for DRs that exploit click feedback: the more extreme the bias is, the lower the DRs' effectiveness. 
  	On contrary, the effectiveness of CoRocchio is left relatively unchanged as $\eta$ increases.
  	This means that, if no click noise is present, CoRocchio can correctly remove the user position bias, no matter how extreme this bias is.
  	On the other hand, for the noise and bias click setting, both Rocchio and CoRocchio show lower nDCG@10 when $\eta$ is larger, but CoRocchio is always better than Rocchio.
  	
  
  In summary, the empirical results analysed in this section suggest that our methods, devised to exploit the implicit feedback signal from historic click logs in the context of DRs, significantly improve the search effectiveness of the considered DRs, regardless of the presence of noise and bias in the click signal.
  

 \subsection{Unseen Queries and CoRocchio-ANN} \label{res-corocchio-ann}
 Next we investigate the effectiveness of our proposed CoRocchio-ANN that can deal with queries that have not been observed in the historical click log: this is done by searching the top-k most similar queries in the click log to the current query, and exploit their clicked passages' representation. The ANN mechanism guarantees to find the k-closest logged queries. We find that setting $k=3$ worked generally well across different DRs and settings; hence we fix $k=3$ throughout this experiment. We note that $k$ could be further fine-tuned for each different DR model by conducting a grid search on a development dataset; we explicitly do not do this for showing the generalizability of our proposed method.

 Note that the ANN procedure can also be applied to Rocchio, thus obtaining Rocchio-ANN. However, the Rocchio and CoRocchio methods (without the additional ANN strategy) cannot be used in this circumstance. The main results are presented in Table~\ref{unseen-results}.
 
 The baseline methods exhibit the same trends observed on the original TREC DL query set (previous section), except that now the PRF methods for DRs are more consistent in the improvements they provide.
 
 The results obtained when exploiting the historical click log with CoRoccho-ANN exhibits substantial improvements over the baselines. The trends observed for unseen queries are similar to those for logged queries (note that the absolute numbers cannot be compared across the two settings, because they refer to different query sets). The highest effectiveness is measured in the perfect and unbiased user setting has the highest effectiveness.
  When position bias is added to the perfect click behaviour, the effectiveness of DRs with Rocchio-ANN decreases. However, DRs with CoRocchio-ANN achieve the same effectiveness obtained when no position bias was present in the click signal. Similar results are found for noisy clicks, but click noise has more of a negative impact than position bias.

For the methods that exploits historic click log in Table~\ref{table:seen_results}, i.e., Rocchio and CoRocchio, they cannot directly deal with unseen queries because there is no associated click data in the log for them to update representation for unseen queries. However, our propose ANN approach for Rocchio and CoRocchio can deal with unseen queries by searching top-k nearest neighbor queries in the click log and exploits their clicked passages' representation. As the results reported in the Table~\ref{table:unseen_results}, both Rocchio-ANN and CoRocchio-ANN shown large improvement over PRF baselines for all the user settings. Similar to the results for the logged query in Table~\ref{table:seen_results}, the perfect and unbiased user setting has the highest effectiveness for DRs uses both Rocchio-ANN and CoRocchio-ANN. When the position bias is added to perfect click behaviour, the effectiveness of DRs with Rocchio is deceased. However, DRs with CoRocchio-ANN achieve the same effectiveness as the setting without position bias. Similar results are found for noise click settings, where CoRocchio-ANN still can remove the position bias for the logged click data hence improving the effectiveness for DRs. Also similar to logged query results, click noise has more negative impact than position bias, and the worse effectiveness scores are observed with the user setting that both click noise and position bias is presented. Although, all the evaluation scores for unseen queries are lower than that of logged queries under the same experimental setting, the improvement of exploiting click feedback from the nearest neighbor queries in the click log is still impressive.

\section{Conclusions and Outlook} \label{conclusion}
In this paper we investigated how to exploit implicit feedback such as the click signal contained in historic click logs to improve the effectiveness of dense retrievers (DRs). With this respect, we proposed to adapt methods used for PRF with DRs to deal with implicit feedback rather than pseudo relevance feedback. To create and investigate effective methods for exploiting the click signal with DRs, we had to overcome three key challenges. 

Challenge 1 related to the absence of a dataset that is sufficient for the training and evaluation of DRs and at the same time contains adequate click information to support our study. To address this challenge, we adapted evaluation practices from online and counterfactual LTR to datasets used for DR evaluation, such as TREC DL. This resulted in the ability of simulating user clicks on SERPs produced by DRs to collect a historic click log. This allowed us to investigate whether the historic click signal improves DRs effectiveness -- we show that our Rocchio method does indeed effectively exploit the click signal and improves DRs effectiveness.

Related to Challenge 1, we also adapted a generative technique for query simulation to augment the click log with a larger set of queries with known relevance judgements. The availability of augmented queries allowed us to study the effectiveness of methods that exploit the historic click signal on queries that have not been observed beforehand. 

Challenge 2 related to the presence of bias (and specifically position bias) in the click signal and its effect on the effectiveness of DR methods. With this respect we found that this bias does reduce the effectiveness of the proposed Rocchio method for exploiting the implicit feedback from historic clicks. To address this, we then devised the CoRocchio method, which counterfactually de-biases the click signal. Theoretical and empirical analyses demonstrated that CoRocchio can effectively address click bias.

Finally, Challenge 3 related to the fact that our counterfactual technique relies on having observed the current user query among those in the historic click log. This is a strong assumption: e.g., the majority of queries observed by web search engines are new queries. To deal with this challenge, we further refined our method to identify in the historic query log queries that are similar (but not identical) to the one currently at hand using the dense query representations. We found that CoRocchio-ANN is capable to effectively exploit the historic click signals of related queries to improve the dense representation (and thus effectiveness) of the current query, which was not observed in the click log.

Our study is the first that investigates how implicit click signal could be exploited in the context of DRs. We believe our work paves the way for the development of effective methods based on DRs that exploit click signals in an online manner. Code and experimental results are publicly available at \url{https://github.com/ielab/Counterfactual-DR}.


\subsubsection*{Acknowledgements.} This research is partially funded by the Grain Research and Development Corporation project AgAsk (UOQ2003-009RTX). 

\bibliographystyle{ACM-Reference-Format}
\bibliography{sigir2022-counterfactual-DR}

\appendix

\end{document}